\documentclass[11pt,a4paper]{article}
\pdfoutput=1
\usepackage[utf8]{inputenc}
\usepackage{verbatim}
\usepackage{amsthm}
\usepackage{amsmath}
\usepackage{amssymb}
\usepackage{graphicx, subfigure}
\usepackage{esint}
\usepackage[normalem]{ulem}

\usepackage{placeins}%multi-column
\usepackage{siunitx}

\usepackage{jcappub}

\begin{document}

\title{The local value of $H_0$ in an inhomogeneous universe}

\author[a]{I. Odderskov,}
\author[a]{S. M. Koksbang,}
\author[a]{S. Hannestad,}

%\affiliation[a]{Department of Physics and Astronomy \\
%University of Aarhus, DK-8000 Aarhus C, Denmark}

\affiliation[a]{Department of Physics and Astronomy \\ University of Aarhus, Ny Munkegade, Aarhus C, Denmark}

\emailAdd{isho07@phys.au.dk, koksbang@phys.au.dk, sth@phys.au.dk}

\date{\today}

\abstract{
The effects of local inhomogeneities on low redshift $H_0$ determinations are studied by estimating the redshift-distance relation of mock sources in N-body simulations. The results are compared to those obtained using the standard approach based on Hubble's law. The comparison shows a clear tendency for the standard approach to yield lower values of $H_0$ than the approach based on the scheme using light rays. The difference is, however, small. More precisely, it is found that the overall effect of inhomogeneities on the determination of $H_0$ is a small increase in the local estimates of about $0.3\%$ compared to the results obtained with Hubble's law, when based on a typical distribution of supernovae in the redshift range $0.01 < z < 0.1$.
\newline\indent
The overall conclusion of the study is a verification of the results that have earlier been obtained by using Hubble's law: The effects of inhomogeneities on local $H_0$ estimates are not significant enough to make it plausible that differences in high- and low-redshift estimates of $H_0$ are due to small inhomogeneities within the setting of standard cosmology.
}
\maketitle

\section{Introduction}
Overall, the $\Lambda$CDM model is consistent with observations at a remarkable level considering the model's simplicity compared to the real, inhomogeneous universe. However, some observational inconsistencies do exist when observations are interpreted within the Friedmann-Lemaitre-Robertson-Walker (FLRW) regime. One such inconsistency is differences in the Hubble constant $H_0$ when determined by different types of observations. In particular, CMB and BAO observations, pertaining to the large scale universe, yield smaller values of $H_0$ than local observations such as those based on supernova redshift-distance relations \cite{Planck,Carnegie,Riess}. As pointed out in {\em e.g.} \cite{Efstathiou}, the different $H_0$ estimates are not actually inconsistent and might for instance simply be due to systematic effects not properly accounted for. Nevertheless, the tension between the different values of $H_0$ is conspicuous and deserves to be studied further.
\newline\indent  
Observations are primarily interpreted using FLRW predictions. The Universe is, however, not an FLRW universe at an exact level and the spatial inhomogeneities and anisotropies of the real universe may affect especially low-redshift observations and their interpretation. Inhomogeneities may in principle also affect large-scale observations such as those of the CMB. The general expectation is that this is not the case though -- an expectation supported by the studies in {\em e.g.} \cite{CMB_distance, lensing_bias, MikkoSyksy_CMB} (see however also\footnote{Note that it is also in \cite{Durrer_CMB} concluded that an inhomogeneous spacetime can have significant effects on $H_0$ estimates based on CMB observations. The conclusion is however based on an erroneous analysis which is later corrected by the same authors in \cite{CMB_distance}.}
 \cite{Bolejko_CMB}). This implies that the tension between the different $H_0$ estimates may be a result of small scale inhomogeneities primarily affecting the low-redshift estimates while the large scale estimates of $H_0$ can be considered yielding the global value. In order to investigate whether or not this is the case, quantitative theoretical considerations regarding low-redshift estimates of $H_0$ are needed. Several different types of studies of effects of inhomogeneities have been conducted, mostly either based on perturbation theory ({\em e.g.} \cite{umeh, ido2, ido3, VT, Bonvin, Clarkson}) or swiss-cheese and onion models ({\em e.g.} \cite{velocity_point, H0_inhomo, tardis, light_cone, marra,bolejko_cmb,Szybka, lensing_effects, dallas_cheese, Kottler_cheese, onion}). The results from such studies are not always consistent with each other and are hence not conclusive at this point, although there is a tendency for the studies to indicate a negligible effect of inhomogeneities on observations after averaging over many observed sources (unless there is significant backreaction \cite{lightprop_Syksy, lightprop_Syksy2, tardis}). However, studies of effects of inhomogeneities only seldom focus on low-redshift $H_0$ determinations. One of the few studies focused on low-redshift $H_0$ estimates was presented in \cite{Durrer_H0} where it was shown that spatial inhomogeneities may affect $H_0$ estimates at an observationally significant level. In contrast to this, studies based on Newtonian N-body simulations indicate that local inhomogeneities do not significantly affect local $H_0$ estimates and thus that local inhomogeneities cannot explain the tension between local and global determinations of $H_0$ \cite{Wojtak:2013,Turner, Io}. A crucial point with the studies based on Newtonian N-body simulations is that they are usually based on Hubble's law. Real astrophysical observations are based on light rays that have traveled through the inhomogeneous universe to reach us. These light rays are affected by inhomogeneities in terms of {\em e.g.} the actual null-paths, their redshift, and their ray-bundle magnification. In the study presented here, such effects are taken into account when computing redshift-distance relations in Newtonian N-body simulations. The resulting redshift-distance data points are fitted to the background FLRW relation with $H_0$ as the fitting parameter. If the discrepancy between the different types of $H_0$ determinations is due to one type (CMB + BAO) probing the global value while the other (low-redshift) studies yield a local value affected by inhomogeneities, then the best-fit $H_0$ value obtained here should be larger than the N-body background value of $H_0$.

\section{Method for computing redshift-distance relations in N-body simulations at low redshift}\label{subsec:doppler}
In order to study the effects of inhomogeneities on the local measurement of $H_0$, mock observations are constructed by computing redshift-distance data points using results from an N-body simulation. The formalism for computing redshift-distance data points is introduced below while practical aspects regarding the N-body simulation are described in section \ref{sec:scheme}.
\newline\newline
A recipe for going between Newtonian N-body simulations and a relativistic spacetime corresponding to the first order perturbed FLRW metric in the Newtonian gauge was given in \cite{recipe}. By using N-body data based on exact, inhomogeneous solutions to Einstein's equations, it was in \cite{Sofie, Sofie2} shown that the recipe leads to a very precise reproduction of redshift-distance relations in an inhomogeneous universe as long as the anisotropy of the individual structures is low. The particular procedure for computing redshift-distance relations studied in \cite{Sofie2} will be used here. The study in \cite{Sofie2} only included mildly non-linear structures, but in relation to the study presented here, a similar precision of the method has been confirmed for density fractions in the range $0.05\lesssim \frac{\rho}{\rho_{bg}} \lesssim 10$, where $\rho_{bg}$ is the density of the assumed ``background". (The precision of the method for inhomogeneities outside this range has not been studied.) The procedure is therefore expected to yield fairly accurate estimates of redshift-distance relations including effects of inhomogeneities.
\newline\indent
The procedure is only summarized below, and the reader is referred to {\em e.g.} \cite{recipe, Sofie2} or the other references cited in this section for elaborations.
\newline\newline
The observed redshift of a light ray with tangent vector $k^{\alpha}$ is computed following the standard definition $z_{\text{obs}}+1 = \frac{(k^{\alpha}u_{\alpha})_e}{(k^{\alpha}u_{\alpha})_0}$. The subscripts $0$ and $e$ indicate evaluation at the spacetime position of observation and emission respectively such that {\em e.g.} $\left( u^{\alpha}\right)|_0$ is the observer velocity and $\left( u^{\alpha}\right)|_e$ is the velocity of the source. The velocity field is given by $u^{\mu} = \frac{c}{V}(1,v^i)$, where $v^i$ are the velocity components obtained from the N-body simulation and the normalization factor $ V \approx \sqrt{c^2 - a^2(v_r^2+ v_{\theta}^2r^2 + v_{\phi}^2r^2\sin^2(\theta))}$ is computed according to the findings of \cite{Sofie2}.
\newline\indent
The effects of inhomogeneities on the angular diameter distance can be described by the convergence $\kappa$ such that the angular diameter distance along light rays in the N-body simulation is given by $D_A\approx D_{A, bg}(1-\kappa)$. $D_{A,bg}$ denotes the angular diameter distance at $z = z_{\text{obs}}$ according to the FLRW background of the N-body simulation. Once the angular diameter distance is known, the corresponding luminosity distance can be computed using the distance duality relation $D_L = (1+z_{obs})^2 D_A$.
\newline\newline
At linear level, the convergence is divided into several different contributions including the (integrated) Sachs-Wolfe contributions, a gravitational contribution and a Doppler contribution. In general, the two dominant contributions are the gravitational convergence and the Doppler convergence. As shown in \cite{bright_side, bright_side2, Sofie2}, the gravitational contribution to the convergence is, however, irrelevant at low redshifts. Hence, for the purpose of the current study, the angular diameter distance can be computed as $D_A \approx D_{A,bg}(1-\kappa_v)$, where $\kappa_v$ denotes the Doppler convergence. The Doppler convergence can be computed by (see {\em e.g.} \cite{Bonvin_doppler}):
\begin{equation}\label{eq:Doppler}
\begin{split}
\kappa_vc =   \left(1-\frac{1}{a_{,t}(\eta_0-\eta)} \right) n^iu_i, \\
\end{split}
\end{equation}
where $\eta$ denotes conformal time and $u_i$ is the (spatial) velocity of the source. The Einstein convention is used with $i$ running over the spatial indices $1,2,3$. The vector $n^{\mu}$ is the propagation direction unit $4$-vector, {\em i.e.} $n^{\mu} \propto k^{\mu}+\frac{k^{\nu}u_{\nu}}{c^2}u^{\mu}$. It is sufficient to use the background version of $n^{\mu}$ and in that case $n^{i}$ is the unit vector pointing in the direction from the source to the observer. 
\newline\indent
In the expression for the Doppler convergence given above, a term due to the peculiar motion of the observer has been omitted. This is justified by noting that the observer velocity can be estimated by measuring the CMB dipole and hence has no consequences when interpreting real observations. Note also that in equation (\ref{eq:Doppler}), the scale factor etc. are evaluated at the spacetime position corresponding to the {\em observed} redshift $z_{\text{obs}}$, not a background redshift computed with the background metric. 
\newline\indent
When mock redshift-distance data points have been obtained as described above, they are fitted to the expression for the luminosity distance in an FLRW universe,
\begin{align}\label{eq:dLz}
D_L(z) &= \frac{(1+z)c}{H_0}\int_0^z \frac{dz'}{\sqrt{\Omega_{m,0}(1+z')^3+\Omega_{\Lambda,0}}}.
\end{align}
When fitting, $H_0$ is used as fitting parameter while $\Omega_{m,0}$ and $\Omega_{\Lambda,0}$ are fixed to the values used in the N-body simulation.
\newline\newline
Before moving on, it should be noted that the Doppler convergence occurs because of the difference between the observed redshift and the redshift obtained from background computations (see {\em e.g.} \cite{bright_side, bright_side2, Bonvin, Bonvin_doppler, xtra_doppler}). The effects of inhomogeneities studied here are therefore effects of peculiar velocities. The peculiar velocities are gravitationally induced by an inhomogeneous energy density and the effects of the peculiar velocities are thus effects of having a spacetime that deviates from the exactly spatially homogeneous and isotropic FLRW spacetimes. This fact will in the following be emphasized by referring to an ``inhomogeneous velocity field" instead of ``peculiar velocities".
\newline\indent
Note lastly, that the effects of an inhomogeneous velocity field on {\em e.g.} local $H_0$ estimates was also studied in \cite{Kaiser}, where it was argued that the effects can actually be understood without invoking general relativity.

\section{Mock observations}\label{sec:scheme}
This section serves to introduce the construction of the mock observations used for the study. The (mock) observations are based on a Newtonian N-body simulation with cosmological parameters in agreement with the 2013 results from the Planck collaboration \cite{Planck}, {\em i.e.} the cosmological parameters are set according to $\left( \Omega_b,\Omega_{\text{CDM}}\right) = (0.048,0.26)$ and ($h,\sigma_8) = (0.68,0.84)$. The simulation is performed using a modified version of the GADGET-2 code \cite{Springel:2005}, with initial conditions generated using a code written by J. Brandbyge \cite{Brandbyge} based on transfer functions computed using CAMB\footnote{http://camb.info/} \cite{camb}. The simulation is run in a box of side length $\SI{512}{Mpc/h}$, containing $512^3$ dark matter particles. The simulation is initiated at a redshift of $z=50$ and run until present time. 
\newline\indent
Box size, simulation resolution and source distribution on the sky will not be varied here as it was in \cite{Io} shown that such variations have only small effects on the results.
\newline\newline
Velocities and positions of sources as well as positions of observers are obtained by identifying these with halos, which are found using the halo-finder ROCKSTAR \cite{Behroozi:2011}.
\newline\indent
As seen in equation (\ref{eq:Doppler}), the velocity fields of the sources are highly important for computing the Doppler convergence. It was demonstrated in \cite{velocity_point} that the effects of inhomogeneities on observations will depend on the smoothing scales used to obtain {\em e.g.} the velocity field from the discrete N-body data. This issue is here overcome by estimating source velocities as the velocities of bound structures, i.e. halos, in the simulation. Assuming that galaxies follow the motion of their dark matter host halos, this procedure corresponds to probing the velocity field in the same manner as is done in real observations.

\subsection{Redshift distribution of sources}
\begin{figure}
\centering
\subfigure{\hspace*{0.1cm}  
	\includegraphics{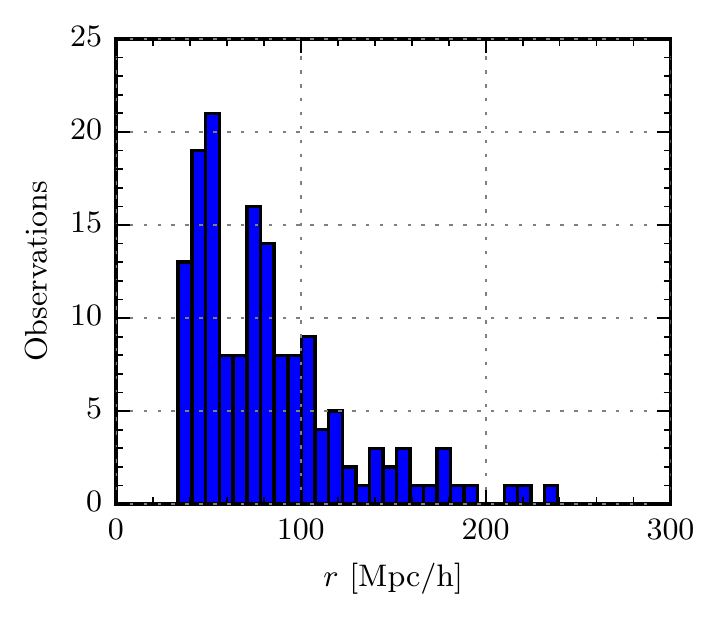}
}
\subfigure{
	\includegraphics{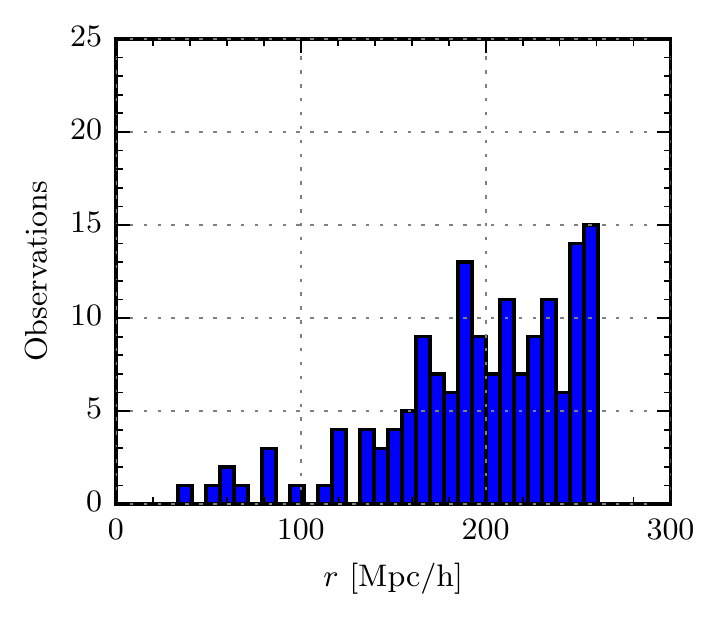}
}
	
\subfigure{\hspace*{-0.4cm}
	\includegraphics{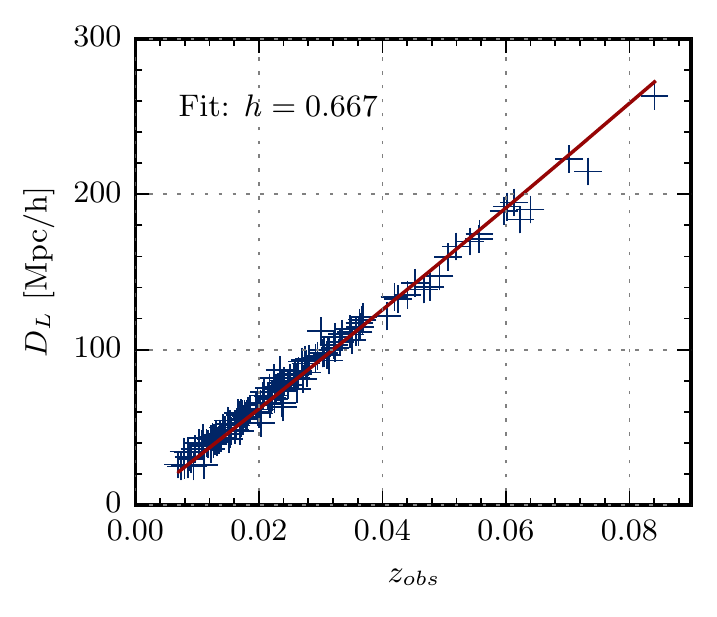}
}
\subfigure{
	\includegraphics{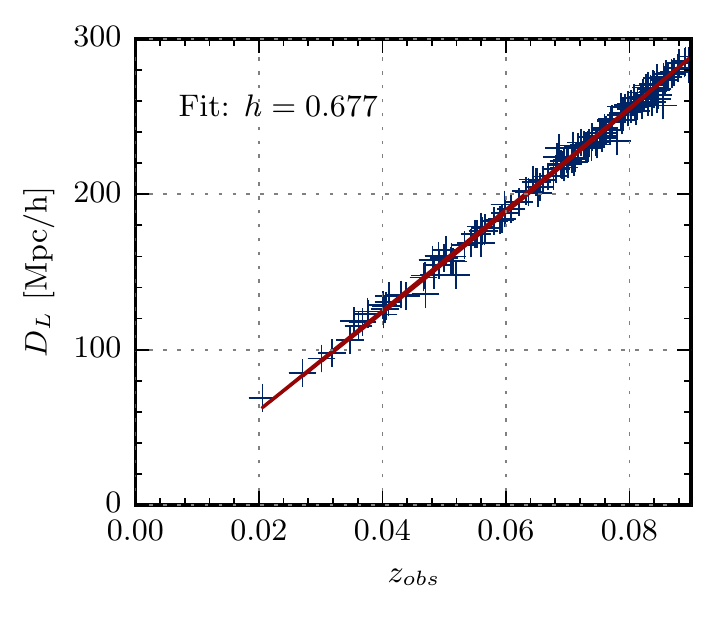}
}

\caption{\textbf{Top}: Distribution of observations as a function of distance. Left: 'Typical' distribution used in estimates of $H_0$ in the local universe, chosen to correspond to the redshift distribution of the 155 SNe of the
CfA3+OLD sample \cite{Jha:2007,Hicken:2009}. Right: Mass weighted distribution. \textbf{Bottom:} Example of fitting mock observations to equation (\ref{eq:dLz}) to obtain $H_0$, given in terms of the dimensionless Hubble parameter $h=H_0/(\SI{100}{km/s/Mpc})$, for the two different redshift distributions.}
\label{fig:redshiftdistributions}
\end{figure}

The redshift distribution of the sources affects the study; the further away the sources are, the less impact the peculiar velocities have. To understand the significance of the sources' redshift distribution, two different redshift distribution schemes are used. In one of the schemes, the sources are distributed according to the redshift distribution of the 155 supernovae of the CfA3+OLD sample \cite{Jha:2007,Hicken:2009}.
The other scheme selects the 155 sources according to a mass weighted distribution, in which each halo in the distance range $\SI{30}{Mpc/h}<r< \SI{256}{Mpc/h}$ is selected with a probability proportional to its mass\footnote{This distribution scheme builds on the assumption that the probability of a type Ia supernova occurring in a given halo is proportional to the halo mass. However, as seen in figure \ref{fig:redshiftdistributions}, the scheme leads to a redshift distribution which is significantly different than the distribution of the CfA3+OLD sample.}. This distribution has the advantage that it is growing as a function of distance (due to the increasing volume of shells). Using this distribution scheme thereby provides a check that the $H_0$ estimates converge to the true value at large distances. Note also that it is this second scheme that was used in \cite{Io}.
\newline\indent
The difference between the two types of distributions is illustrated in figure \ref{fig:redshiftdistributions}. The figure also shows an example of the differences in $H_0$ estimates based on observations following the two schemes.

\subsection{Observers}\label{subsec:peebles_henvisning}
In order to take cosmic variance into account\footnote{The term ``cosmic variance" is here used to mean the uncertainty in observations due to our ability to only observe the Universe from a single spacetime position.}, each set of mock observations are based on a group of observers instead of a single observer. The observer positions are still important for the study as the local environment may bias observations at low and intermediate redshifts. For instance, placing the observers in massive halos amounts to placing them in infall regions and will thus lead to a bias of peculiar source velocities, and hence redshifts, in the direction towards the observer which will again lead to a smaller $H_0$ estimate. This is predicted by gravitational instability theory \cite[Chapter~5]{Peebles}, and can in principle be corrected for by subtracting the peculiar motion induced by the local density field when analyzing real observational data (see {\em e.g.} \cite{Neill}). Such a correction was for instance made in \cite{Riess}.
\newline\indent
In order to study the effects of the local environment of the observers, two different sets of observers are used. The first set consists of observers residing in subhalos of mass $10^{12}-10^{13} M_\odot/h$ in a host halo in the mass range $5\cdot 10^{14}-5\cdot 10^{15} M_\odot/h$. This approximately corresponds to our position in the Local Group galaxy cluster, a member of the Virgo Super Cluster. The second set of observers are placed at random positions throughout the simulation volume. As underdense regions take up the larger part of the volume, these observers will tend to be positioned in underdense regions.
\newline\indent
When observers are placed in halos, their peculiar velocities are taken to be those of the host halos. When they are placed at random positions, they are assumed to be at rest with respect to the background. The observer velocities enter into the computations of the observed redshift and so using these two different methods for estimating observer velocities may affect the results. This has been tested not to be the case though; setting the observer velocities to zero in the case where the observers are placed in halos is insignificant for the results.

\subsection{Lightcone snapshots} \label{subsec:lightcone}
The velocity in equation (\ref{eq:Doppler}) is the comoving velocity of the source at the time of emission. The mock observations are carried out in the halo catalog at $z=0$. However, comoving source positions change over time due to the sources' peculiar velocities (which also change over time). Since the study made here is based on low-redshift observations with $z\lesssim 0.1$, this cosmic evolution should not affect the results at a significant level. To test this assessment, two types of halo catalogs are made. One is based on a regular GADGET-2 snapshot at $z = 0$, while the other type is based on lightcone snapshots\footnote{The lightcone snapshots are created by using a plug-in to GADGET-2 written by Troels Haugbølle. The same plug-in was used in \cite{Io}.} (one for each observer). ROCKSTAR's lightcone functionality can be used for identifying structures in such a snapshot. In the resulting halo catalog, the positions and velocities are stored as they were at the time of emission.

\subsection{Summary}\label{subsec:mock_observations}
\begin{figure}
\centering
\includegraphics{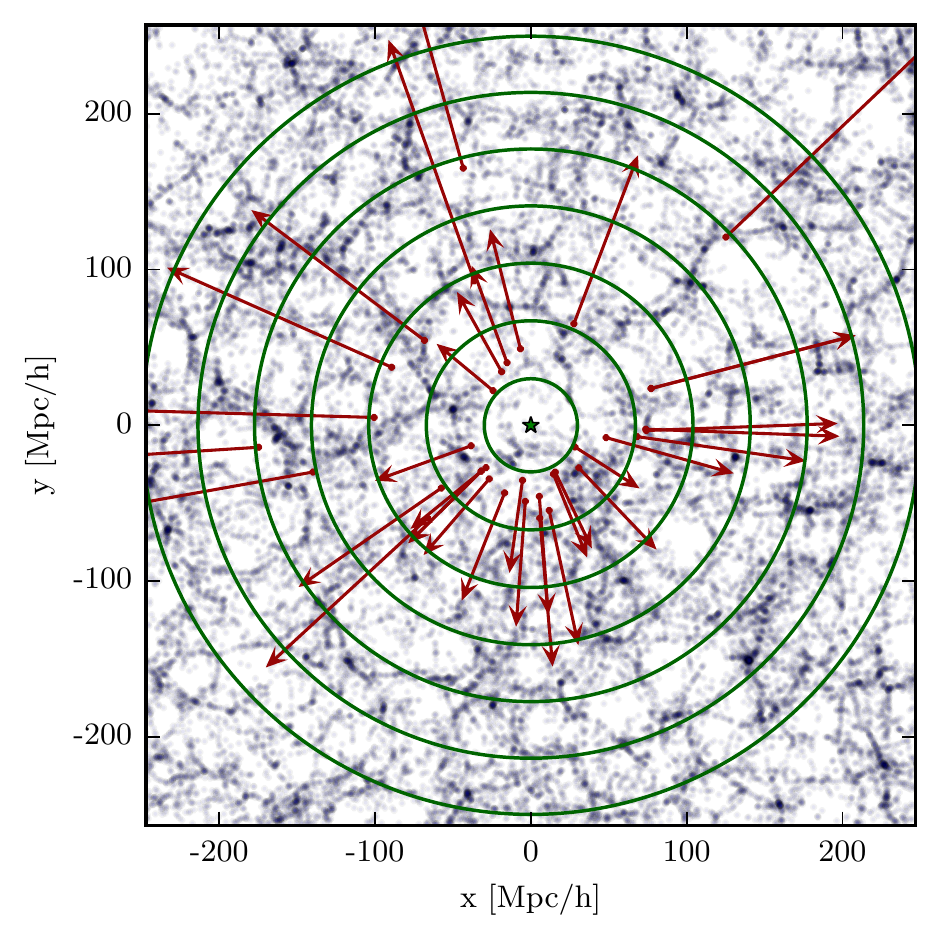}
\caption{Example of the observations carried out by one of the observers (indicated by the green star). The observed halos are colored red, and their total physical velocities relative to the observer are shown with red arrows. The division of redshift bins is illustrated with green circles -- only every 5th division is shown.}
\label{fig:Illustration_of_observations}
\end{figure}

In figure \ref{fig:Illustration_of_observations}, an example of the observations performed by a single observer is shown. As seen, the observers view the full sky. See {\em e.g.} \cite{Io} for the effects of using smaller sky coverage.

Following the various schemes discussed above, a number of different analyses have been performed. These are summarized in 
table \ref{tab:analyses}.
\newline\indent
Since the lightcone snapshots are very time consuming to use, only a single analysis is made with these. That analysis is denoted Dlc and is the analysis which best resembles real observations.

\begin{center}
\begin{table}
\center %
\begin{tabular}{clcccc}
\hline\hline
 Figure & Analysis name & Observer positions & SN-distribution & Fit & Lightcone\\
\hline 
% &  &  &  & \tabularnewline
%   &     & \textbf{Figure \ref{fig:Nbody_observations1}} & & \\
  & Dref & Subhalos         & Typical       & Full   & No\\
\ref{fig:Nbody_observations1} & Href & Subhalos         & Typical       & Linear & No\\
%   &     & \textbf{Figure \ref{fig:Nbody_observations2}:} & & \\
&&&& \\
  & Dlc  & Subhalos         & Typical       & Full   & Yes \\
  & Dmw  & Subhalos         & Mass Weighted & Full   & No\\
\ref{fig:Nbody_observations2} & Drp  & Random Positions & Typical       & Full   & No\\
  & Hrp  & Random Positions & Typical       & Linear & No\\
  & Hmw  & Subhalos         & Mass Weighted & Linear & No\\ 
\hline
\end{tabular}\caption{Overview of analyses included in the study. The analysis names are given such that analyses with $H_0$ estimates based on Hubble's law ("linear" fit) start with an "H" while those with $H_0$ estimates based on computing $D_L$ using the Doppler convergence and fitting to the full $D_{L,\Lambda\text{CDM}}$ expression ("full" fit) start with a "D". The last part of the name denotes how each analysis deviates from the reference ("ref") analysis, respectively by using lightcone snapshots ("lc"), using  a mass weighted selection of SNe ("mw"), and placing the observers at random positions instead of in halos ("rp").}
\label{tab:analyses}
\end{table}
\end{center}

\FloatBarrier

\section{Results and discussion}

\begin{figure}[t]
\centering
% % % % % % % % % %
\subfigure
	{
	\includegraphics{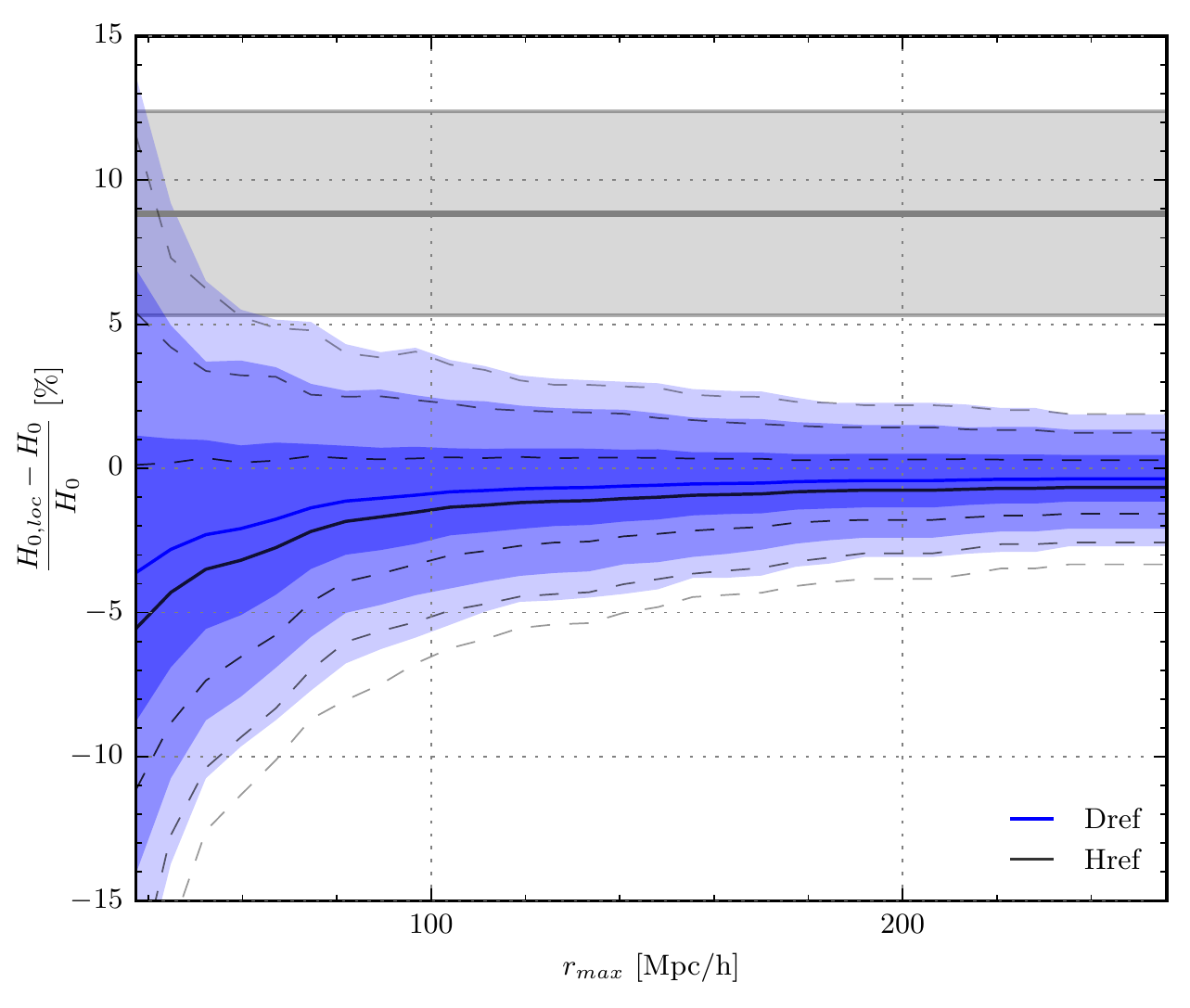}
	}
% % % % % % % % % %
\caption{Deviations between $H_0$ of the N-body background and the local values $H_{0,loc}$ of $H_0$ obtained using the analyses Dref and Href described in table \ref{tab:analyses}. Shadings and dashed lines indicate 68.3\%, 95.4\% and 99.7\% confidence intervals while solid lines indicate mean values. The results are shown as a function of the maximum distance $r_{\max}$ between the observer and observed objects included in the $H_0$ estimates. The gray area indicates the observed value and uncertainty of $H_0$ found in \cite{Riess}.} 
\label{fig:Nbody_observations1}
\end{figure}

\begin{figure}[t]
\centering
% % % % % % % % % %
\subfigure
	{
	\includegraphics{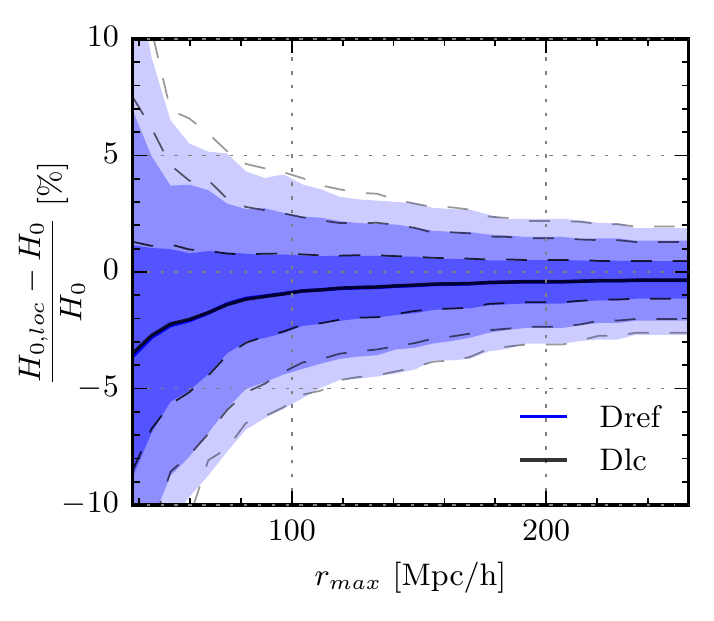}
	\includegraphics{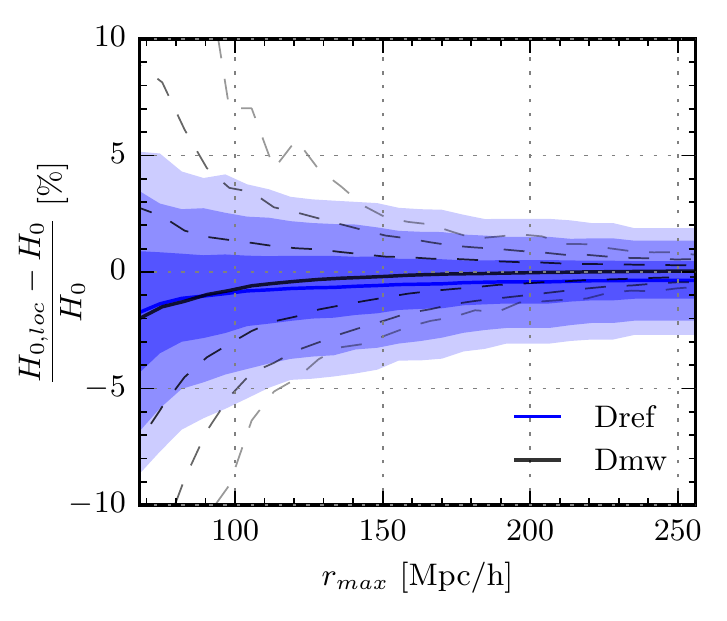}
	}	
% % % % % % % % % %
\subfigure
	{
	\includegraphics{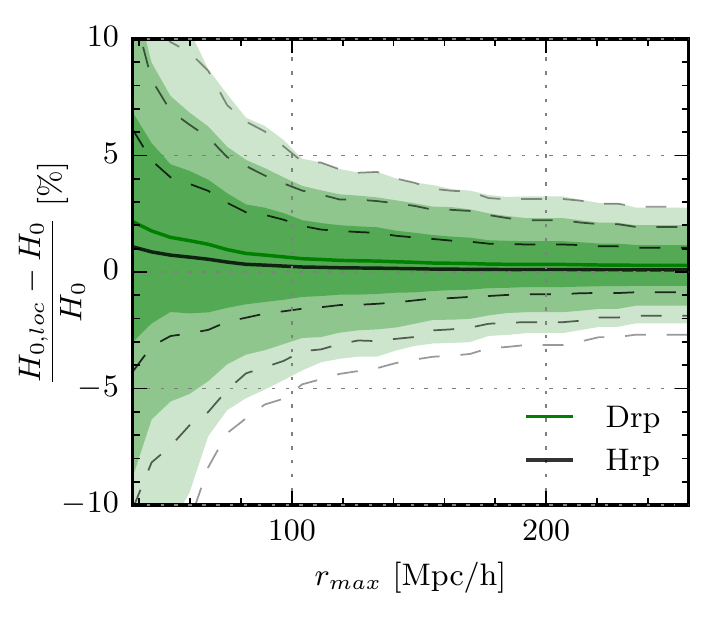}
	\includegraphics{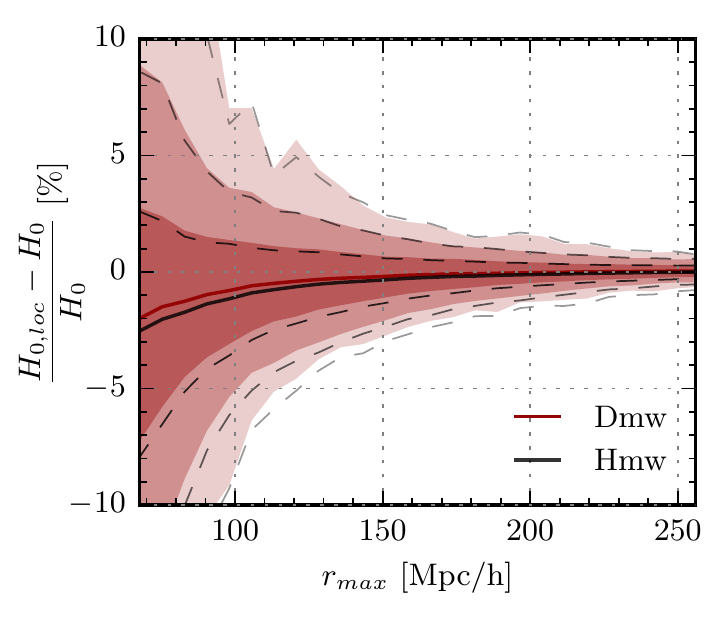}
	}
\caption{Deviations between $H_0$ of the N-body background and the local values of $H_0$ obtained using the analyses listed in table \ref{tab:analyses} as indicated in the figure legends. Shadings and dashed lines indicate 68.3\%, 95.4\% and 99.7\% confidence intervals while solid lines indicate mean values. The results are shown as a function of the maximum distance $r_{max}$ between the observer and observed objects.
}
\label{fig:Nbody_observations2}
\end{figure}

Figures \ref{fig:Nbody_observations1} and \ref{fig:Nbody_observations2} show how the local estimates of $H_0$ are distributed among the observers in the different analyses. Specifically, the deviation in the local Hubble constant from the true value, $(H_{0,loc}-H_0)/H_0$, is shown as a function of the distance $r_{max}$ to the most distant bin included in the estimate of $H_{0,loc}$. The sources are distributed in bins according to their comoving distances from the observer, which are related to the redshifts of the FLRW background via $r = \frac{c}{H_0}\int_0^z \frac{dz'}{\sqrt{\Omega_{m,0}(1+z')^3+\Omega_{\Lambda,0}}}$.
The mean value of the deviation is shown with a solid line, and the 68.3\%, 95.4\% and 99.7\% confidence intervals are indicated with either shaded areas or dashed lines. 
\newline\newline

The results are shown using different combinations of choices for the observer positions, redshift distribution of supernovae, and fitting method. An overview of the different combinations that have been studied is given in table \ref{tab:analyses} where a naming scheme for the analyses is also introduced. Analysis Dlc is the analysis which is closest to actual observations; it uses observers placed in subhalos with a typical supernova distribution in a lightcone snapshot and obtains $H_0$ by a fit to the FLRW redshift-distance relation as explained in section \ref{sec:scheme}. In figure \ref{fig:Nbody_observations2}, analysis Dlc is compared to analysis Dref which differs from Dlc only in that it is based on a regular GADGET-2 snapshot. As seen, the difference between using the two types of snapshots is negligible.
\newline\indent
Figure \ref{fig:Nbody_observations1} compares analysis Dref to analysis Href. Analysis Href differs from analysis Dref in that $H_0$ has been obtained using Hubble's law. The comparison shows that the mean $H_0$ estimate is smaller when Hubble's law is used instead of the recipe based on the Doppler convergence as described in section \ref{subsec:doppler}. The difference is quite small, ranging from 1.9\% in the innermost bin to 0.3\% in the outermost bin. A similar result is seen in figure \ref{fig:Nbody_observations2}, {\em i.e.} the analyses based on Hubble's law consistently yield a $H_0$ estimate that is slightly smaller than the estimate of the corresponding analysis based on the Doppler convergence computations described in section \ref{subsec:doppler}.
\newline\newline

Figure \ref{fig:Nbody_observations2} illustrates the importance of using realistic source and observer distributions. In particular, the figure shows that by changing the observer positions to be random, the $H_0$ estimates grow. This is because underdense regions take up more of the simulation volume and hence will be favored when using random observer positions. The result can thus be considered a type of Hubble-bubble effect (see {\em e.g.} \cite{bubble1,bubble2, bubble4, bubble5, bubble6, bubble8, bubble9}). This type of Hubble-bubble result is particularly interesting since there in fact are some indications that we may be living in an underdense region of the Universe \cite{local_structure1, local_structure2, local_structure3}.
\newline\indent
The figure also shows that the distribution of sources has impact on especially the uncertainty of the $H_0$ estimates. Specifically, the spread in the results is much less when higher-redshift sources are favored in the observations. This is consistent with peculiar motions being less important at larger distances.
\newline\newline
Lastly, it is noted that the results presented here yield mean $H_0$ estimates that are {\em lower} than the $H_0$ value of the N-body background. The only exception to this result is the analyses with random observer positions discussed above. Within the standard cosmological scheme studied here, the effect of peculiar velocity fields thus seems to be to {\em reduce} the local value of $H_0$ compared to the global value. Hence, inhomogeneities cannot provide a solution to the tension of different types of $H_0$ estimates. In fact, the results show that inhomogeneities will in principle increase the tension. As mentioned in section \ref{subsec:peebles_henvisning}, to the extent that the shift in the mean estimate of $H_0$ compared to the global value is a result of the local environment at the observer position, this can be corrected for when interpreting real observations.

\FloatBarrier

\section{Summary}
N-body simulations were used to study low-redshift determinations of $H_0$ in an inhomogeneous universe. In particular, it was studied whether or not an inhomogeneous velocity field leads to a local $H_0$ value different from the global value, the latter corresponding to that of the N-body background and which is what one expects to measure using high-redshift observations such as those based on the CMB. The $H_0$ values were obtained by approximating light paths according to the background model and computing the angular diameter distance of objects by using the Doppler convergence to take effects of inhomogeneities into account. Inhomogeneities were also taken into account when computing source redshifts. This procedure differs from that used in earlier work of this kind, where $H_0$ determinations have been estimated by using the linear relation known as Hubble's law. The results obtained with the method introduced here consistently lead to a mean $H_0$ value which is higher than that obtained by simply using Hubble's law. The difference is, however, very small.
\newline\indent
The $H_0$ estimates were found to be slightly lower than the background value except in a single of the studied cases: When the observers are placed at random positions in the simulation box, the resulting mean $H_0$ estimate is slightly above that of the N-body background. This result is similar to the Hubble-bubble effect.
\newline\indent
As source distances and numbers are increased, all the $H_0$ estimates converge towards the value of $H_0$ of the N-body background.
\newline\newline
The results summarized above indicate that the tension between high- and low-redshift determinations of $H_0$ cannot be alleviated by taking inhomogeneities into account -- at least not without straying from standard cosmology.

\section{Acknowledgments}
We thank the anonymous referee for suggestions that have significantly improved the presentation of the work.
\newline\indent
Parts of the numerical work for this project have been done using computer resources from the Center for Scientific Computing Aarhus.

\end{document}